\documentclass[sigconf]{acmart}

\usepackage{multirow}
\usepackage{enumerate}
\usepackage{float}
\usepackage{dblfloatfix}
	
\usepackage{soul}

\AtBeginDocument{%
  \providecommand\BibTeX{{%
    \normalfont B\kern-0.5em{\scshape i\kern-0.25em b}\kern-0.8em\TeX}}}

\setcopyright{acmcopyright}
\copyrightyear{2022} 
\acmYear{2022} 
\acmConference[EuroUSEC 2022]{2022 European Symposium on Usable Security}{September 29--30, 2022}{Karlsruhe, Germany}
\acmPrice{15.00}
\acmISBN{978-1-4503-9700-1/22/09}

\begin{document}

\title[ENAGRAM: An App to Evaluate Preventative Nudges for Instagram]{ENAGRAM: An App to Evaluate Preventative Nudges\\ for Instagram} 

\author{Nicol\'{a}s E. D\'{i}az Ferreyra}
\affiliation{%
  \institution{Hamburg University of Technology}
  \city{Hamburg}
  \country{Germany}
  }
\email{nicolas.diaz-ferreyra@tuhh.de}

\author{Sina Ostendorf}
\affiliation{%
  \institution{University of Duisburg–Essen}
  \city{Duisburg}
  \country{Germany}
  }
\email{sina.ostendorf@uni-due.de}

\author{Esma A\"{i}meur}
\affiliation{%
  \institution{University of Montr\'{e}al}
  \city{Montr\'{e}al}
  \country{Canada}
  }
\email{aimeur@iro.umontreal.ca}

\author{Maritta Heisel}
\affiliation{%
  \institution{University of Duisburg-Essen}
  \city{Duisburg}
  \country{Germany}
  }
\email{maritta.heisel@uni-due.de}

\author{Matthias Brand}
\affiliation{%
  \institution{University of Duisburg-Essen}
  \city{Duisburg}
  \country{Germany}
  }
\email{matthias.brand@uni-due.de}

\renewcommand{\shortauthors}{D\'{i}az Ferreyra et al.}

\begin{abstract} 
Online self-disclosure is perhaps one of the last decade's most studied communication processes, thanks to the introduction of Online Social Networks (OSNs) like Facebook. Self-disclosure research has contributed significantly to the design of \textit{preventative nudges} seeking to support and guide users when revealing private information in OSNs. Still, assessing the effectiveness of these solutions is often challenging since changing or modifying the choice architecture of OSN platforms is practically unfeasible. In turn, the effectiveness of numerous nudging designs is supported primarily by self-reported data instead of actual behavioral information. \textbf{Objective}: This work presents ENAGRAM, an app for evaluating preventative nudges, and reports the first results of an empirical study conducted with it. Such a study aims to showcase how the app (and the data collected with it) can be leveraged to assess the effectiveness of a particular nudging approach. \textbf{Method}: We used ENAGRAM as a vehicle to test a risk-based strategy for nudging the self-disclosure decisions of Instagram users. For this, we created two variations of the same nudge (i.e., with and without risk information) and tested it in a between-subjects experimental setting. Study participants (N=22) were recruited via Prolific and asked to use the app regularly for 7 days. An online survey was distributed at the end of the experiment to measure some privacy-related constructs. \textbf{Results}: From the data collected with ENAGRAM, we observed lower (though non-significant) self-disclosure levels when applying risk-based interventions. The constructs measured with the survey were not significant either, except for participants' External Information Privacy Concerns (EIPC). \textbf{Implications}: Our results suggest that (i) ENAGRAM is a suitable alternative for conducting longitudinal experiments in a privacy-friendly way, and (ii) it provides a flexible framework for the evaluation of a broad spectrum of nudging solutions.
\end{abstract}



\maketitle

\section{Introduction} \label{sec:introduction}
Over the last decades, the use of information and communication technologies has widely extended across different segments of everyday life. From making bank transactions to finding a life partner, online services have acquired an increasingly important role in the dynamics of modern societies \citep{royakkers2018societal}. Nevertheless, living in a digitalized world also introduces threats and challenges related to privacy and security since online services are fed and operate over large amounts of personal data. At the same time, technology must create adequate cybersecurity conditions to safeguard the privacy rights and the integrity of its users. For this, legal frameworks —such as the EU General Data Protection Regulation (GDPR)— were introduced to enforce tech companies to comply with a set of data protection principles. Overall, this contributes to ensure a secure processing and storage of personal data, preventing its non-consensual exploitation, and avoiding unfair discrimination of data subjects, among others \citep{de2018consent}. However, to a large extent, the privacy decisions and practices of individuals have also been challenged by the affordances of media and communication technologies \citep{azzouz2019strategies}. Particularly, Online Social Networks (OSNs) like Facebook or Instagram have redefined and blurred people’s privacy boundaries by creating spaces in which they can connect seamlessly and share personal information with large (and sometimes untrusted) audiences~\citep{kramer2019mastering}.

Like in the real world, individuals disclose private information in OSNs to create and maintain social relationships with others \citep{penni2017future,wang2016modeling}. Thereby, the strength of such relationships tends to increase, and so does people's social capital \citep{ellison2011connection}. However, deciding whether or not to disclose personal information to others is not always straightforward (even more so in online environments) \citep{kramer2019mastering}. To a large extent, this may because online self-disclosure decisions are likely driven by short-term gratifications (e.g., likes, comments, or number of followers) instead of long-term privacy risks \citep{ostendorf2020neglecting}. In turn, users are prone to disclose sensitive information unseemly to untrusted recipients and becoming victims of privacy threats such as reputation damage, social engineering, and even financial fraud \citep{aimeur2018scourge}. Moreover, OSNs often hinder individuals' self-presentation decisions as they place different audiences (e.g., work colleagues and family) in a common communication plane \citep{vitak2012impact}. Consequently, users frequently experience regret —along with unwanted incidents— after sharing personal information with unintended recipients \citep{wang2011regretted}.

All in all, interaction in OSNs can lead to unwanted incidents even on platforms exhibiting secure backend infrastructures and compliant with data protection principles \citep{albladi2016vulnerability,de2018consent}. To a great extent, information and cues about the potential risks of unrestrained self-disclosures can help users regulate their exposure levels and mitigate their chances of experiencing negative consequences \citep{diaz2020preventative,gerber2019investigating,Ostendorf2022Cyber}. Nonetheless, current layouts and graphical interfaces of OSNs do not provide any means (i.e., cues or information) that may help users determine the potential risks and hazardous outcomes of their disclosures \citep{terpstra2019improving}. Conversely, platforms showcase many cues not only related to immediate gratification (e.g., like buttons) but also related to their reputation (e.g., their size) or recognition (e.g., their market presence) that contribute to larger self-disclosure levels, since such cues increase trust in the platform, and larger self-disclosure levels, in turn, serve their business model. In some cases, individuals can even develop problematic/addictive usage behaviors. Although more research is needed, a problematic/addictive usage of OSNs may (among other factors) arise due to application-specific features and affordances that especially enable the experience of immediate gratification while potentially hindering self-control and further reflective processes \cite{brand2022can}. Moreover, the privacy policies of OSNs are also devoid of information related to potential privacy threats and leave rational risk estimations to the individual discretion of each user \citep{terpstra2019improving,de2018consent}. Hence, there is an urgent call for technological affordances that promote safe and more reflective information-sharing decisions among the users of OSNs, so that the risks of online-self disclosure are mitigated or prevented.

\subsection*{Motivation}

Over recent years, privacy scholars have introduced a wide range of technological approaches that aim to improve people's online privacy decisions \cite{mosca2020agent,misra2017pacman,sanchez2018co,salem2020,schobel2020understanding}. In particular, the use of nudges has gained popularity due to their capacity for assisting and guiding individuals towards safer privacy practices \cite{acquisti2017nudges}. At their core, nudges are interventions that encourage a certain behaviour which, in turn, tends to maximize people's welfare \cite{ThalerSunstein08}. Such interventions are the means for behavioural change and consist of small modifications in the context within which decisions are made \cite{lin2017nudge}. For instance, displaying cues related to the targeted audience of a post can motivate users to employ custom friend lists \cite{wang2011regretted}. Given the close relation existing between risk perception and privacy behaviour, it is not surprising that interventions portraying risk information are deemed adequate for motivating safer self-disclosure decisions in OSNs \cite{marmion2017cognitive,samat2017formatvscontent,kramer2019mastering}. In particular, such interventions can prevent users from sharing posts with personal data by rendering information about the risks of unsafe self-disclosure practices \citep{gerber2019investigating}.

Despite researchers’ increasing interest in nudges and their applications to cyber-security, many solutions have remained theoretical or at early design stages (c.f., \cite{salem2020,diaz2020preventative,kroll2021digital,schobel2020understanding}). This is because nudges are frequently conceived as interventions that should be integrated into preexisting choice architectures. That is, into current OSNs platforms or services. However, the most popular platforms do not offer integration mechanisms that would allow researchers to assess nudges’ effectiveness within their intended operational environment. In turn, many approaches receive partial evaluation from end-users through mock-ups and self-reports. Hence, there is a call for evaluation approaches in which preventative nudges can be tested and assessed under more realistic working conditions.



\subsection*{Contribution}

In this work we present ENAGRAM, an app to evaluate preventative nudges for Instagram. ENAGRAM is an independent 3rd-party Android application that acts as an Instagram proxy. As with the original Instagram app, it allows users to elaborate posts using free text and pictures (e.g., photos taken with their phones). However, it also incorporates a nudging mechanism consisting of interventions (or pop-up messages) that are triggered at the moment of sharing the posts. Users' reactions to these interventions (i.e., whether they ignored them or not) along with some supplementary information (e.g., a hashed version of the post) are recorded by the app in a dedicated server for later analysis.

We conducted a pilot study via Prolific (N=22) to showcase ENAGRAM's evaluation features and functionalities. For this, we used the app to test a risk-based strategy in which information about common OSN incidents (e.g., reputation damage, identity theft, etc.) is applied to nudge users towards safer self-disclosure decisions. We chose this particular nudging approach as a running example since it is part of our prior work (see \cite{diaz2020preventative}). Nevertheless, ENAGRAM's interventions can be tailored for assessing other nudging solutions alike. All in all, the contribution of this paper is two-fold: (i) it presents and discusses ENAGRAM's affordances for the evaluation of preventative nudges, and (ii) reports the results of a preliminary study on the effectiveness of the implemented nudging strategy.



The remainder of this paper is organized as follows. In the next section we provide the paper's background and discuss related work concerning the design and evaluation of preventative nudges for OSNs. Next, in Section~\ref{sec:architecture} we introduce ENAGRAM's main architectural components, whereas in Section~\ref{expe_design} we describe the methodology applied for its empirical assessment (7-day between-subjects approach). The results of our preliminary study are presented in Section~\ref{sec:results} and then discussed in Section~\ref{sec:discussion}. Particularly, we analyze ENAGRAM's benefits and drawbacks based on a set of constructs and performance metrics collected by the end of the experiment. Finally, we present the limitations of our approach in Section~\ref{sec:limitations} and conclude with some prospective directions for future work in Section~\ref{sec:conclusion}.

\section{Background and Related Work}

Since its introduction by Nobel Prize winners Richard Thaler and Cass Sunstein, the ``nudge'' theory has been extensively investigated and applied repeatedly to the design of privacy-enhancing technologies. When it comes to OSNs, there is a wide variety of privacy nudges in the current literature whose goal is to support users' online self-disclosure decisions \citep{icissp21,botti2021automatic,diaz2020preventative,masaki2020exploring,salem2020, raber2016privacy}. For instance, \citet{raber2016privacy} introduced \textit{PrivacyWedges}, a visualization strategy for aiding the audience selection of social media publications. Under the premise that ``close'' contacts are often the most trustworthy ones, \textit{PrivacyWedges} displays network members based on their interpersonal distance to the targeted user (i.e., well-known friends are prioritized over the rest). Thereby, the user is encouraged to keep sensitive posts within her inner circle of friends and thus away from unintended recipients. Other nudging solutions precisely seek to warn users about publications containing sensitive information. Such is the case of \citet{botti2021automatic}, who applied a compound of Natural Language Processing (NLP) techniques for determining whether a post contains information about one's location, health, and personal identifiers, among others. In line with this, \citet{icissp21} empirically assessed users' perception of warnings about the presence of personal information in their posts. All in all, most study participants considered such nudges useful but were agnostic regarding their future adoption.

\citet{wang2013privacy} conducted perhaps one of the most groundbreaking studies on the use of privacy nudges in OSNs. They implemented three different nudges that intervened when users were about to post something on Facebook: (i) an \textit{audience} nudge showed visual cues about the potential recipients of the post (i.e., pictures of friends), (ii) a \textit{sentiment} nudge displayed the sentiment of the text being posted, and (iii) a \textit{timing} nudge delayed the actual publication of the post for some minutes. These interventions were designed to give users the chance to re-think their disclosures, edit, or even withdraw them before publication. The findings of this experiment not only yielded valuable evidence on the effectiveness of preventative nudges but also served as a reference for later contributions elaborating on the same (or closely-related) intervention strategies. Such is the case of \citet{masaki2020exploring}, who incorporate information about frequent online privacy harms to the nudge's design in a very similar way. Other solutions like the ones of \citet{diaz2020preventative} and \citet{salem2020} also stress the importance of personalization to increase nudges' efficacy. Particularly, they seek to overcome the limitations of one-size-fits-all approaches by adjusting the frequency and content of interventions to the individual privacy goals of each user.

Despite the great attention these nudges have received in recent years, only a reduced number of the proposed solutions have been implemented and tested under realistic conditions. That is, through evaluation instruments other than mock-ups and self-reports. As pointed out by \citet{gomez2018experiments}, ``\textit{...experiments are but a few when compared to surveys and theoretical approaches, even when adopting a broad definition of experiment}''. When looking closer at some of the few implemented solutions, it is clear that integration is a major challenge as none of them is heavily embedded into a commercial OSN platform (e.g., Facebook, Twitter, or Instagram). For instance, the nudges of \citet{wang2013privacy} were implemented as browser extensions scrapping Facebook's web interface, whereas a recent approach by \citet{alemany2022} was developed and tested in a non-commercial platform. In the case of browser extensions, one limitation relates to the evolving nature of OSNs, which makes them obsolete after significant changes on the site's interface are made. Furthermore, as users gravitate toward mobile apps, browser-based experimental conditions may become less engaging for research participants and thus unfeasible for conducting longitudinal studies. On the other hand, while non-commercial OSNs offer suitable nudge integration means, they may be alien to most study participants and thus fail to recreate interaction patterns like the ones emerging within commercial sites. Therefore, there is a need for evaluation methods and tools that help researchers overcome said integration limitations of proprietary OSNs.

\section{ENAGRAM Architecture} \label{sec:architecture}

\begin{figure*}[t]
    \centering
    \includegraphics[height=8cm]{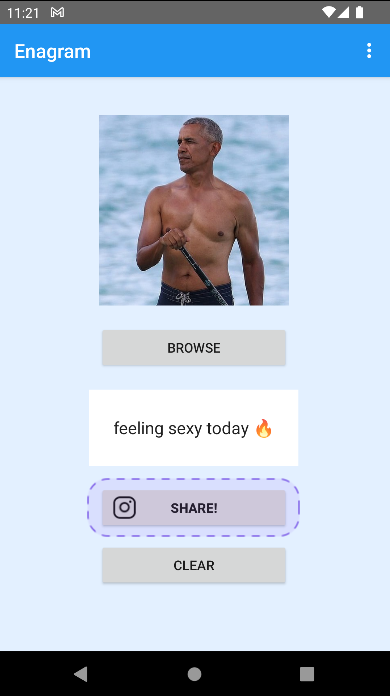}\hspace{10pt}
    \includegraphics[height=8cm]{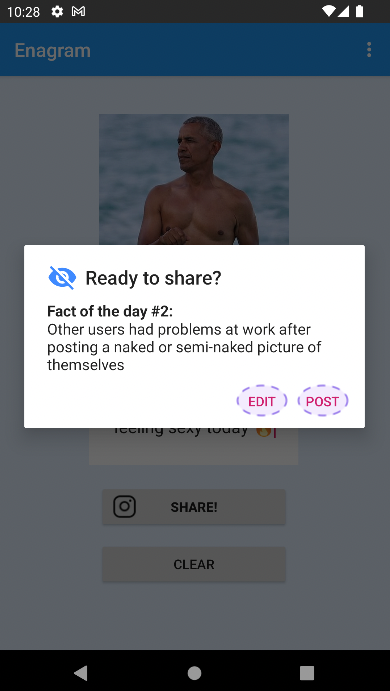}\hspace{10pt}
    \includegraphics[height=8cm]{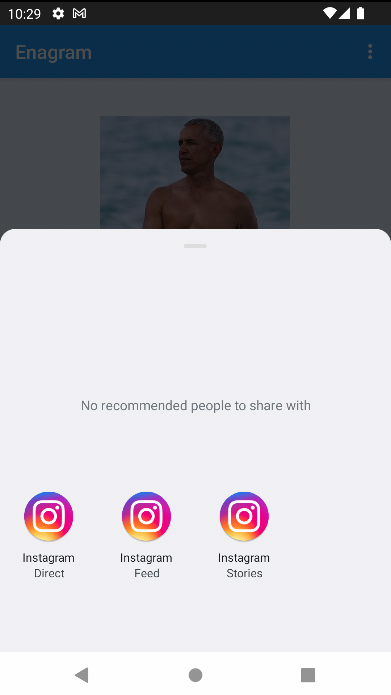}
    \caption{ENAGRAM interfaces for Group 2: Post composition (a), Risk-based intervention (b), and posting selection (c).}
    \label{fig:screenshots}
\end{figure*}

ENAGRAM is an Android app created to support the research and evaluation of preventative nudges on Instagram. It incorporates features for capturing users' reactions towards behavioral interventions while offering a flexible framework for testing different variants of such interventions.

As shown in \autoref{fig:screenshots}-a, ENAGRAM provides the basic functionality for creating Instagram publications, namely picture selection and caption composition features. At its core, ENAGRAM acts as an Instagram proxy: it forwards the post information (i.e., photo and caption text) to the Instagram app via Android intents. However, before passing the control over to Instagram, ENAGRAM intervenes with a pop-up message. Such a pop-up can, for instance, display the legend ``Ready to share?'' and a ``fact of the day'' describing a self-disclosure privacy threat (\autoref{fig:screenshots}-b). Users then have the chance to proceed and post their messages on Instagram or to go back to the composition screen (\autoref{fig:screenshots}-a) and edit them. If a user decides to continue, she can choose whether to post the message on her Instagram feed, create a story, or send it via direct message (\autoref{fig:screenshots}-c). This nudging strategy proposed by \citet{diaz2020preventative} aims to promote reflective self-disclosure practices among OSNs users by means of risk information and awareness. However, ENAGRAM could be extended/tailored for testing other nudging solutions alike (see Section~\ref{sec:benefits}).

\autoref{fig:architecture} (left) illustrates ENAGRAM's main architectural components. It follows a REpresentational State Transfer (REST) communication schema where client (Android app) and server (HTML web server) interact via HTTP methods (e.g., GET, POST, and DELETE) using a lightweight data-interchange format (i.e., JSON). Communication is done through an Application Programming Interface (API) consisting of a collection of methods and operations such as \textit{Login}, \textit{Logout}, \textit{Register}, and \textit{Post}. Such API methods are PHP implemented and allow retrieving (pushing) content from (to) an SQL database (\texttt{EventsDB}) containing information about users' interactions within ENAGRAM. As shown in \autoref{fig:architecture} (right), the \texttt{EventsDB} consists of 5 tables:
\begin{enumerate}[i]
    \item \textbf{\texttt{users\_table}}: Contains the login credentials (username and password) of registered users, their current app version, and app language. As we describe later in Section~\ref{expe_design}, we created two versions of ENAGRAM, both in English and German (i.e., 4 variants in total).
    \item \textbf{\texttt{interventions}}: Contains the intervention messages shown within the app and some additional information (e.g., a risk value) that could be leveraged to adapt the display frequency of the corresponding privacy prompts (c.f.,~ \cite{diaz2020preventative}). ENAGRAM includes a total of 26 different intervention messages.
    \item \textbf{\texttt{intervention\_categories}}: Intervention messages are grouped around 6 large categories (i) \textit{drugs and alcohol use}, (ii) \textit{sex}, (iii) \textit{religion and politics}, (iv) \textit{strong sentiment}, (v) \textit{location}, and (vi) \textit{personal identifiers}. Hence, each entry in the \texttt{interventions} table belongs to one of these categories (e.g., the message displayed in \autoref{fig:screenshots}-b belongs to the ``sex'' category). Both intervention messages and categories correspond to the ones curated~by~\citet{diaz2020preventative}.
    \item \textbf{\texttt{popup\_actions}}: Defines the type of actions a user can take when interacting with the nudge pop-up:
\begin{itemize}
\item \textit{action\_id} = 0: The user clicked ``edit'' after receiving an intervention.
\item \textit{action\_id} = 1: The user clicked ``post'' after receiving an intervention.
\end{itemize}
To minimize the chances of habituation biases and annoyance, the current version of ENAGRAM intervenes at most 5 times a day with a time interval of 60 min. between interventions. Hence, it may happen that a user may not receive an intervention after clicking on ``SHARE!'' (e.g., if she wishes to share a picture 10 min. after being nudged for the first time). To keep track of all the sharing events within the app, we included the following action type:  
\begin{itemize}
\item \textit{action\_id} = 2: The user clicked ``SHARE!'' in the main window but did not receive an intervention afterwards.
\end{itemize}

\begin{figure*}[!t]
    \centering
    \includegraphics[height=3.5cm]{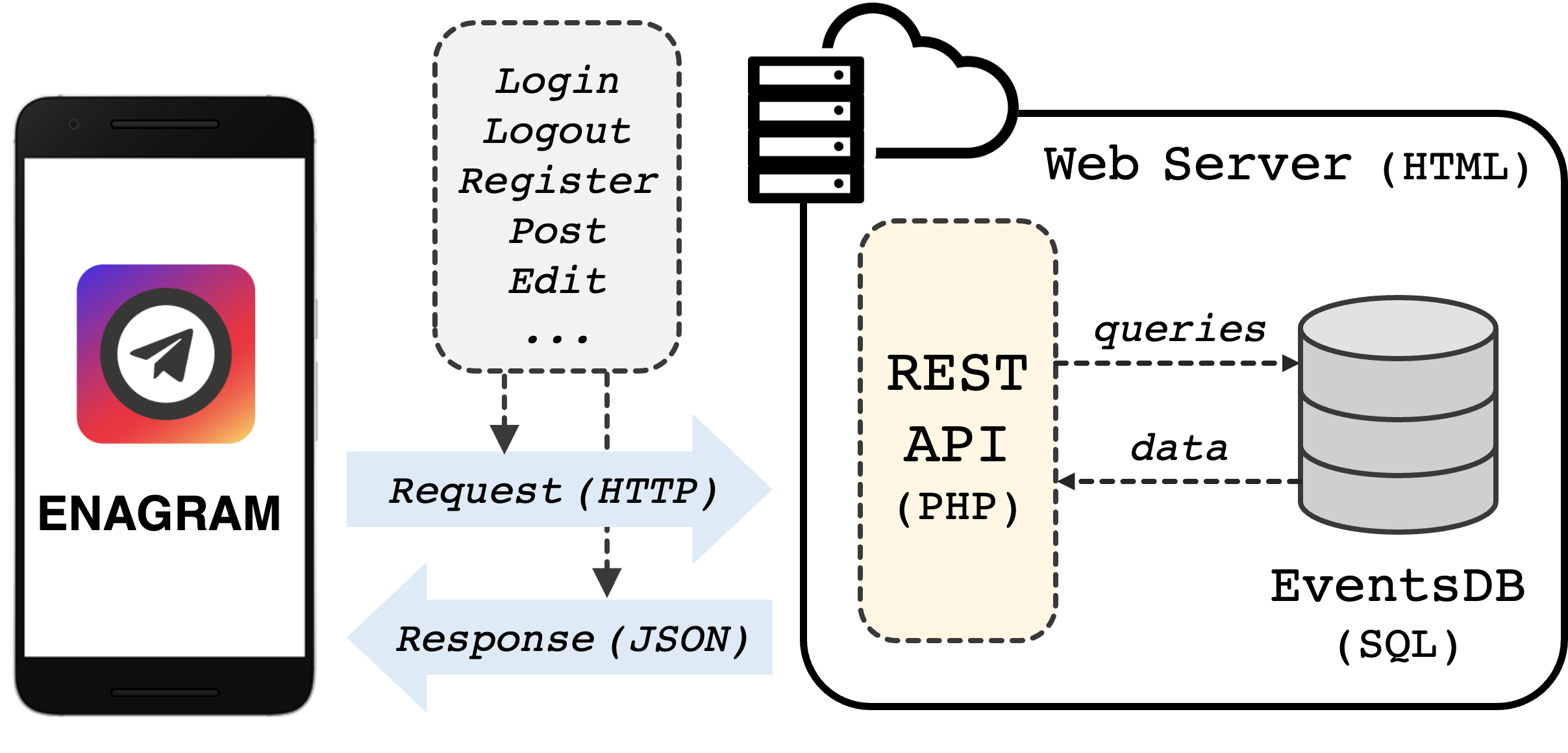}
    \hspace{10pt}
    \includegraphics[height=3.5cm]{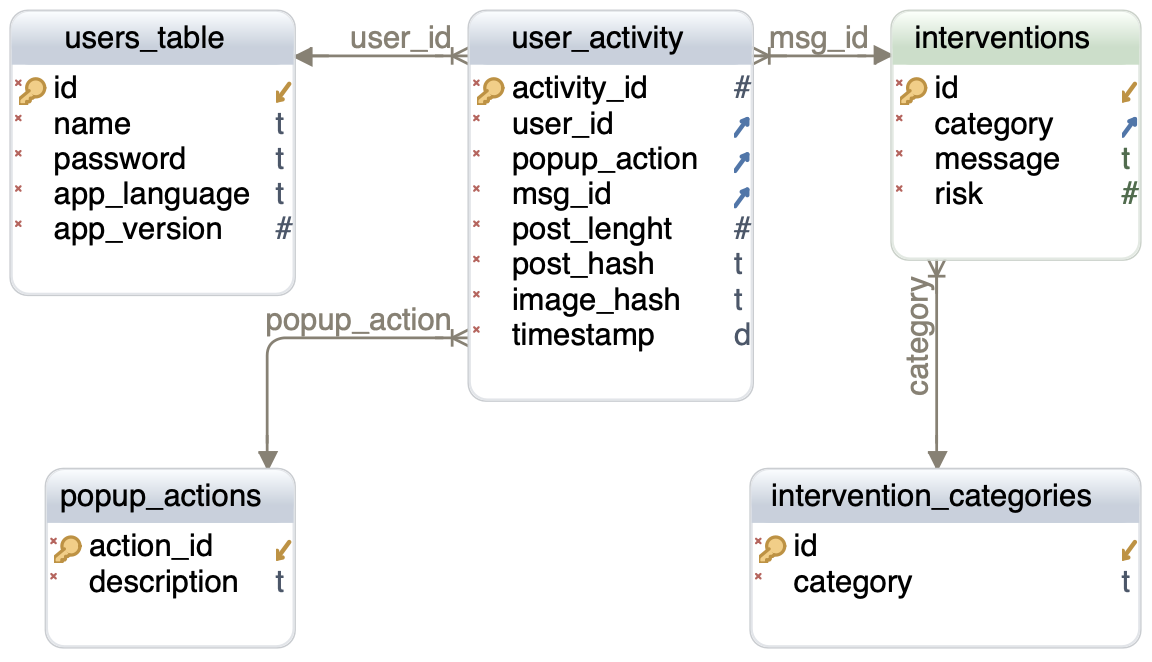}
    \caption{ENAGRAM architecture (left) and EventsDB schema (right).}
    \label{fig:architecture}
\end{figure*}

\begin{figure*}[!b]
    \centering
    \includegraphics[width=0.85\linewidth]{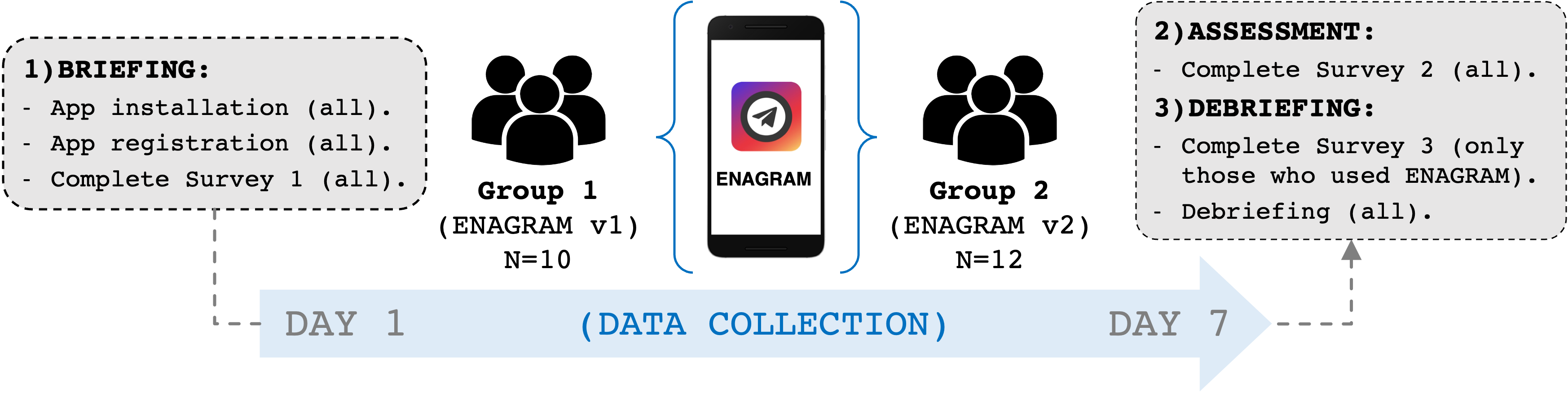}\vspace{-2ex}
    \caption{Study design workflow (group sizes refer to the number of participants who completed the experiment).}
    \label{fig:methodology}
\end{figure*}

\item \textbf{\texttt{user\_activity}}: This table contains all the events recorded by ENAGRAM for all its users. Such events correspond to the actions listed in the \texttt{popup\_actions} table characterized by the following contextual information:

\begin{itemize}
\item \textit{popup\_action}: The type of action being recorded (i.e., 0, 1, or~2).
\item \textit{user\_id}: The id of the user whose action is being recorded.
\item \textit{msg\_id}: A number between 1 and 26 pointing to the id of the warning message being displayed in the pop-up. This field is \texttt{null} when \textit{popup\_action} is equal to 2.
\item \textit{post\_lenght}: Number of characters in the picture caption.
\item \textit{post\_hash}: A hashed version (i.e., a pseudonym consisting of a fixed-size sequence of hexadecimal characters) of the picture caption. This can be used to check whether the user changed the caption after receiving an intervention.
\item \textit{image\_hash}: A hashed version of the picture file path. Like the previous one, it can be used to check whether the user selected a different picture after receiving an intervention.
\item \textit{timestamp}: The time at which the event occurred.
\end{itemize}
The \texttt{user\_activity} table is populated every time the user clicks on ``SHARE!'', ``EDIT'', or ``POST''. Hence, it can be seen as a collection of snapshots describing the user's self-disclosure behavior over time.
\end{enumerate}

\section{Experimental Design} \label{expe_design}

We conducted a preliminary study on a particular nudging approach to explore ENAGRAM's evaluation features. As mentioned in Section~\ref{sec:introduction}, we decided to test the strategy proposed in \cite{diaz2020preventative} since it is part of our prior work. Thus, the outcome of this study provides (i) actionable information about the effectiveness of such a strategy and (ii) empirical evidence about ENAGRAM's benefits and drawbacks. Whereas the remaining of this paper focus mainly on the former point, the latter is discussed thoroughly in Section~\ref{sec:benefits}.

We created 2 versions of the app and tested them in a between-groups experimental setting. The two versions only differed in the intervention pop-up: \textit{version 1} (v1) only displayed the legend ``Ready to share?'' (i.e., without showing any risk information) and \textit{version 2} (v2) included also the ``fact of the day''. As mentioned in Section~\ref{sec:architecture}, users were nudged at most 5 times a day with a minimum gap of 60 min. between interventions. In addition, v2 users did not receive the same threat description twice in the same day.

\subsubsection*{Recruitment} 

The study participants were recruited via Prolific\footnote{\url{https://prolific.co}} and assigned to one of the two experimental conditions: participants in Group 1 tested ENAGRAM v1 and participants of Group 2 tested ENAGRAM v2. The group assignment was done pseudo-randomly to create a gender balance within each experimental condition. Participants had to be active Instagram users, at least 18 years old, and had to have an Android phone with OS version 9.0 or higher (API 28). We asked them to have at least 1GB free space in their devices and the latest Instagram version installed. Having a computer/laptop/notebook was also a requirement as we included some questionnaires as part of the study.

\subsubsection*{Study approach} 

The study consisted of three subsequent stages, namely \textit{briefing}, \textit{assessment}, and \textit{debriefing} (\autoref{fig:methodology}). During the \textit{briefing}, participants were asked to install ENAGRAM on their phones and answer some demographic questions (e.g., gender, age, and average time spent on Instagram). We used the following \textit{cover story} to avoid behavioral biases: participants were told that ENAGRAM was built following a software development method created by university researchers, and that their job was to test the app for \textbf{7 days} and report possible implementation flaws (e.g., glitches, errors, missing requirements). As part of the briefing, each participant received a short tutorial describing the installation steps and instructions for creating an ENAGRAM username and password. A registration code was generated by ENAGRAM which participants had to provide to show they actually installed the app. We explicitly asked them to use a pseudonym as username to track their performance during the whole study. Good command of the German language was also required as we tested the app on its German version.

After 7 days we conducted an \textit{assessment} and \textit{debriefing} of the study participants. The \textit{assessment} consisted of a sort survey asking the participants if they used ENAGRAM regularly in the last 7 days or not. Those who reported not having used the app were then debriefed and fully informed about the actual purpose of the study (viz., test an app containing a nudge mechanism for online self-disclosure). Otherwise, they were asked to complete another survey containing questions about the \textit{performance} of the app and on the following privacy-related constructs: \textit{perceived risks} (RSK), \textit{perceived control} (CTRL), \textit{perceived benefits} (BEN), and \textit{external information privacy concerns} (EIPC). All constructs were previously elaborated and validated by other authors (i.e., EIPC by \citet{morlok2016sharing} and the rest by \citet{Krasnova2009}) and measured using a 7-point Likert scale ranging from 1 = ``strongly disagree'' to 7 = ``strongly agree''. A summary of all employed constructs can be found in the Appendix.

\begin{figure*}[!b]
    \centering
    \includegraphics[height=6cm]{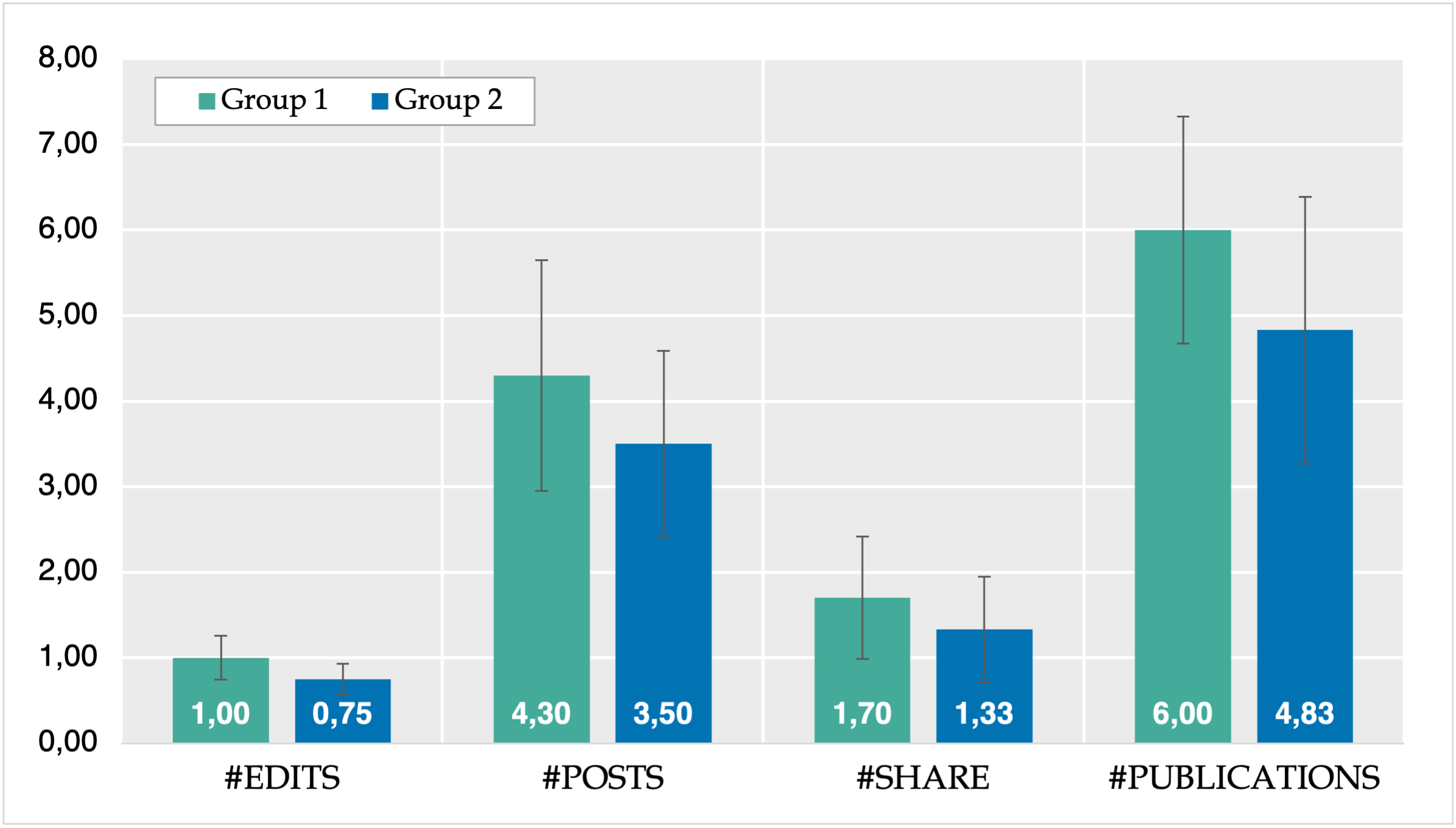}
        \caption{Average number of events recorded by the end of the experiment per group.}
    \label{fig:events}
\end{figure*}

\subsubsection*{Ethical considerations} 
The study was conducted in accordance with the Declaration of Helsinki and approved by the local Ethics Committee of the Department of Computer Science and Applied Cognitive Science of the University of Duisburg-Essen. All participants received information about the study procedure (including data privacy statements) and were asked to give their informed consent before moving forward in the different experimental stages. They could withdraw at any time receiving a compensation for each completed part of the study: 6€ after the briefing, another €6 after the assessment, plus €5 for the final performance questionnaire. In all cases (i.e., after withdrawing or completing the study) participants were debriefed accordingly and then asked whether we could still use their data for research purposes. Survey instruments, software, and study results are available as \hyperref[sec:supplementary]{\textbf{Supplementary Material}}.

\section{Results} \label{sec:results}

We recruited 24 participants for the study (12 for each group condition). One participant from Group 1 was discarded after the assessment stage (not having used the app) and another one by the end of the study (not having answered the control questions correctly). Hence, we considered and analyzed the data gathered from 22 subjects: Group 1 consisted of five females, four males, and one non-binary person (19-34 years, \emph{M} = 23.30, \emph{SD} = 5.27), and Group 2 of eight females and four males (18-33 years, \emph{M} = 23.33, \emph{SD} = 5.40).

\subsection{Behavioral insights} \label{sec:behav_in}

We conducted a manual inspection of the data collected by ENAGRAM stored in the \texttt{EventsDB}. After a sanity check, we observed some duplicated entries in the \textit{user\_activity} table probably due to connection issues on the client side (e.g., the participant's phone lost connection at the moment of sending the data to ENAGRAM' web server). Such duplicated entries were removed resulting on a collection of 137 events: 19 edits (\#EDITS) and 118 publications (\#PUBLICATIONS) from which 85 (\#POSTS) correspond to post actions performed after an intervention (i.e., within the intervention pop-up) and 33 to share actions that were not followed by an intervention (\#SHARES). All in all, we registered 53 interventions across all participants in Group 1 and 51 across all participants in Group 2.

Edit events are of special interest for measuring the effectiveness of the nudging approach. Particularly, cases in which users change the picture or the caption after receiving an intervention could help us determine whether the nudge has indeed an impact on privacy behavior. Such cases can be identified through the \textit{post\_length}, \textit{post\_hash}, and \textit{image\_hash} values of EDIT events (\textit{action\_id}=0) that are closely followed (i.e., regarding \textit{timestamp}) by SHARE! events (\textit{action\_id}=2). If any of these values change from one event to the other, then such a change can be interpreted as an effect of the nudge on the user's self-disclosure behavior. Nevertheless, we identified only 2 cases in which a participant changed either the picture or the caption after clicking on edit (one from Group 1 and the other from Group 2). The rest of the editing cases did not include any changes neither in the selected picture nor in the caption.

\subsection{Group comparisons}

\autoref{fig:events} shows the average \#EDITS, \#POSTS, \#SHARES, and \#PUBLICATIONS per study group. As one can observe, all of these values are higher in Group 1 than in Group 2. To determine whether such differences are statistically significant, we conducted an independent sample \textit{t}-Test (\autoref{table:ttest_descriptives}). Since Levene's test for equality of variances was non-significant in all cases ($p>0.05$), the corresponding $t$ statistics were computed assuming homogeneity of variances. Overall, we found no significant differences between any of the values obtained for Group 1 and Group 2. The effect sizes we obtained were in general ``small'' according to Cohen's convention~\cite{cohen1997}.

We repeated this analysis with the constructs elicited by the end of the experiment (i.e., RSK, CTRL, BEN, EPIC). From \autoref{table:group_descriptives}, we can see that, with the exception of CTRL, all construct values are higher for Group 2 than for Group 1. Once again, we assumed homogeneity of variances when conducting the \textit{t}-Tests as Levene's test was non-significant in all cases ($p>0.05$). As shown in \autoref{table:ttest_descriptives}, the differences between the values obtained for Group 1 and Group 2 were only significant for the EPIC construct. For the rest of the constructs, such differences were not statistically significant. These results yielded ``large'' effect sizes for CTRL and EIPC, ``medium'' for BEN, and ``small'' in the case of RSK.

\begin{table*}[t]
\def\arraystretch{1.2}
\centering
\caption{Descriptive group statistics.}
\small
\begin{tabular}{ p{5.7cm} c c c c c } 
\toprule  
\multicolumn{1}{c}{\textbf{Dependent Variable}} &
\multicolumn{1}{c}{\textbf{Group}} &
\multicolumn{1}{c}{\textbf{N}} &
\multicolumn{1}{c}{\textbf{Mean}} &
\multicolumn{1}{c}{\textbf{SD}} &
\multicolumn{1}{c}{\textbf{SE}} \\ \bottomrule
\multirow{2}{*}{Number of ``edits'' after intervention (\textbf{\#EDITS})} 
    &1 & 10 & 1.000 & 0.816 & 0.258 \\
    &2 & 12 & 0.750 & 0.622 & 0.179 \\ \hline
\multirow{2}{*}{Number of ``post'' after intervention (\textbf{\#POSTS})} 
    &1 & 10 & {4.300} & {4.270} & {1.350}  \\
    &2 & 12 & {3.500} & {3.778} & {1.091}  \\ \hline
\multirow{2}{*}{Total number of ``shares'' (\textbf{\#SHARES})} 
    &1 & 10 & {1.700} & {2.263} & {0.716} \\
    &2 & 12 & {1.333} & {2.146} & {0.620} \\ \hline
\multirow{2}{*}{Total number of ``publications'' (\textbf{\#PUBLICATIONS})} 
    &1 & 10 & 6.000 & 4.190 & 1.325 \\
    &2 & 12 & 4.833 & 5.408 & 1.561 \\ \hline
\multirow{2}{*}{Perceived Risk (\textbf{RSK})} 
    &1 & 10 & 4.025 & 1.003 &  0.317 \\
    &2 & 12 & 4.396 & 1.281 &  0.370 \\ \hline
\multirow{2}{*}{Perceived Control (\textbf{CTRL})} 
    &1 & 10 & 4.667 & 1.432 & 0.453  \\
    &2 & 12 & 3.389 & 1.441 & 0.416  \\ \hline
\multirow{2}{*}{Perceived Benefits (\textbf{BEN})} 
    &1 & 10 & 4.850 & 0.727 & 0.230 \\
    &2 & 12 & 5.313 & 0.765 & 0.221 \\ \hline
\multirow{2}{*}{External Information Privacy Concerns (\textbf{EIPC})} 
    &1 & 10 & 2.900 & 1.233 & 0.390 \\
    &2 & 12 & 4.111 & 1.072 & 0.309 \\
  \bottomrule
\end{tabular}
\label{table:group_descriptives}
\end{table*}\unskip

\begin{table*}[t]
\def\arraystretch{1.2}
\centering
\caption{Independent samples t-Test.}
\small
\begin{tabular}{ r c c l c c c c} 
\toprule  
\multicolumn{1}{c}{\textbf{Dependent Variable}} &
\multicolumn{1}{c}{\textbf{t}} &
\multicolumn{1}{c}{\textbf{d.f.}} &
\multicolumn{1}{c}{\textbf{Sig.}} &
\multicolumn{1}{c}{\textbf{Mean diff.}} &
\multicolumn{1}{c}{\textbf{SE\textsubscript{DM}}} &
\multicolumn{1}{c}{\textbf{95\% CI}} &
\multicolumn{1}{c}{{\textbf{Cohen's d}}}
\\ \bottomrule
\textbf{\#EDITS} &
   0.816 & 20 & 0.424 & 0.250 & 0.307 & (-0.389,0.889) & {0.345}\\
\textbf{\#POSTS} &
   {0.466} & 20 & {0.646} & {0.800} & {1.716} & {(-2.779, 4.379)} & {0.280}\\
\textbf{\#SHARES} &
   0.389 & 20 & 0.701 & 0.367 & 0.942 & (-1.598, 2.331) & {0.168}\\
\textbf{\#PUBLICATIONS} &
   0.556 & 20 & 0.584 & 1.167 & 2.097 & (-3.207, 5.541) & {0.241}\\
\textbf{RSK} &
   -0.744 & 20 & 0.466 & -0.371 & 0.499 & (-1.411, 0.669) & {-0.322}\\
\textbf{CTRL} &
   2.077 & 20 & 0.051 & 1.278 & 0.615 & (-0.006, 2.561) & {0.890}\\
\textbf{BEN} &
   -1.444 & 20 & 0.164 & -0.463 & 0.320 & (-1.131, 0.206) & {-0.620}\\
\textbf{EIPC} &
   -2.466 & 20 & 0.023\textsuperscript{*} & -1.211 & 0.491 & (-2.236, -0.187) & {-1.048}\\
  \bottomrule
\multicolumn{8}{c}{\footnotesize \underline{Note}: (*) The~mean difference is significant for $\alpha=5\%$.}
\end{tabular}
\label{table:ttest_descriptives}
\end{table*}

\section{Discussion}\label{sec:discussion}

The results of this preliminary study provide not only insights about the effects of risk-based interventions, but also on the benefits (and limitations) of ENAGRAM when evaluating preventative nudges. Based on our experience, we discuss the implications of the analysis presented in Section~\ref{sec:results} along with some important aspects that should be taken into consideration when conducting studies using ENAGRAM. Particularly, with regard to (i) the instrumentation of self-disclosure metrics and constructs, and (ii) the technical benefits and drawbacks of the app. 



\subsection{Self-Disclosure Metrics and Constructs} \label{sec:discussion_nudge}

As mentioned in \autoref{sec:behav_in}, four self-disclosure metrics were considered for this study: \#EDITS, \#POSTS, \#SHARES, and \#PUBLICATIONS. Nevertheless, several other metrics could be elaborated with the data collected through ENEGRAM. For instance, each of these metrics could be expressed per time unit (e.g., per day or per week) or in a relative way (e.g., \#EDITS/\#PUBLICATIONS or \#EDITS/\#POSTS). The upper limit of interventions (5 per day in this case) could also be leveraged for the elaboration of metrics. That is, by dividing the number of interventions a user received by the end of the experiment (i.e., \#EDITS + \#POSTS) over the maximum number of interventions the app can generate within the experimental period (7 days x 5 interventions/day = 35 interventions).

Due to the small amount of data collected in the 7 days of experiment, we decided to analyze the self-disclosure behavior of the study participants only through absolute metrics. We observed that participants in Group 1 interacted more with the app than the ones in Group 2 (\autoref{fig:events}). Prior work has emphasized the role that risk cues play in the self-disclosure behavior of OSNs users (c.f., \cite{gerber2019investigating,samat2017formatvscontent}). Particularly, that users' perceived risk of information sharing is one of the most important factors influencing such a behavior. Hence, the risk information displayed on the second version of the app may have (i) lessened the sharing frequency of the study participants within that group, and (ii) increased their perception of privacy risks (\#RSK) at the end of the experiment (i.e., at day 7). Still, none of these differences were found significant and thus should be further investigated and analyzed.

Differences in participants' perceived control (CTRL), benefits (BEN), and external information privacy concerns (EIPC) were also observed at the end of the study. Regarding the former, our results differ from the ones of \citet{kroll2021digital} who observed higher (though marginal) levels of perceived control on those users aware of the presence of privacy nudges in OSNs. However, the nudges tested in such a study did not render any risk information, which may be determining for users' perception of control \cite{hajli2016exploring}. On the other hand, a vast amount of literature has emphasized that higher levels of risk awareness can negatively impact the perceived benefits of online self-disclosure and increase users' privacy concerns \cite{ostendorf2020neglecting,kramer2019mastering,hajli2016exploring}. Hence, a lower BEN and a higher EPIC among Group 2 participants can be also related to the presence of risk information in the corresponding app interventions. This is particularly interesting as EIPC encompasses \textit{social} privacy concerns towards other users, especially with regard to organizational practices affecting other people's privacy. Such concerns play an important role in OSN platforms like Instagram since users can easily compromise the privacy of other individuals when sharing group pictures. Hence, it is to expect that those with higher EIPC would be more reluctant to share information or pictures portraying others \cite{morlok2016sharing}. Nevertheless, our results are still preliminary and call for additional research efforts.



\subsection{ENAGRAM's Benefits and Drawbacks} \label{sec:benefits}

\subsubsection*{Extensibility} Overall, the behavioral data collected through ENAGRAM helped us to gain insight in the self-disclosure practices of the study participants. In this particular case, we took the approach proposed by \citet{diaz2020preventative} and embedded it into the app for its evaluation. However, other nudging strategies (e.g., social norms, pop-out policies, or defaults \cite{caraban201923}) could also be easily implemented as many of ENAGRAM's building blocks can be customized with just a few lines of Java code. Such is the case of the time spent between interventions or the content displayed within them. For the latter, additional changes in the \texttt{EventsDB} may be necessary, particularly in the \texttt{interventions} table as it contains the text placed inside ENAGRAM's pop-up window. Other graphical elements displayed in this window (e.g., the legends ``Ready to share?'' and ``Fact of the day \#N'') can also be adjusted and adapted to the specific needs of each intervention strategy. Furthermore, users' reactions to ENAGRAM's interventions could be leveraged to achieve personalization. That is, by regulating the frequency and the content of each warning according to the number of \textit{edits} and \textit{shares} performed by each user in a given time frame~\mbox{\cite{diaz2020preventative}}.


\subsubsection*{Pending Features} Some aspects and functionalities of ENAGRAM still require further development. Such is the case of the interventions displayed by the app which, at the moment, are not content-dependent. Hence, a user sharing a picture (or caption) about her drinking beer may not necessarily receive a warning message referring to alcohol consumption (e.g., ``Other users had problems at work after posting about their alcohol consumption''). This issue could be addressed by integrating a machine learning solution capable of classifying the content being disclosed by the user (i.e., picture, caption, or both). There is prior and ongoing research in this realm that could be leveraged for this purpose (e.g., \cite{botti2021automatic,ferwerda2018you,song2018picture}) and even commercial platforms offering services for the automatic classification of multimedia content (e.g., Microsoft Azure Computer Vision\footnote{\url{https://azure.microsoft.com/en-us/services/cognitive-services/computer-vision/}} and Google Vision AI\footnote{\url{https://cloud.google.com/vision/}}). Hence, content-aware interventions could be (in principle) shaped by integrating such off-the-shelve solutions into ENAGRAM's architecture. Still, this may not be a straight forward task as the integration of third party software often demands changes and adaptations in the targeted architecture to overcome compatibility issues.

\subsubsection*{Users' Privacy} As shown in Section~\ref{sec:behav_in}, the information collected by ENAGRAM (e.g, the post length and the hashed version of the image path) is useful to spot changes in participants' self-disclosure behavior while preserving their privacy. In principle, such information is enough to (i) identify changes in text or (ii) determine whether a picture has been replaced after an EDIT event. However, it is insufficient to determine whether such changes entail more or less information self-disclosure. For instance, a post like ``I live in New York City'' is shorter but more precise than another one saying ``I live in the United States of America''. Likewise, two different picture paths can tell us that both images are different but not if one is more or less sensitive than the other. Methods like the ones proposed in the previous point can address this issue by collecting metadata (e.g., text sentiment, named entities, or picture explicitness) from the content disclosed within ENAGRAM. That is, by pre-processing the posts (i.e., text and image) and storing the corresponding metadata in the EventsDB for later analysis. Thereby, the effects of ENAGRAM's interventions could be better assessed without having to record participants' raw data.

\subsubsection*{Technical Issues} Participants also had the chance to report any technical problem they may have experienced while using the app. Some of them mentioned that the caption was not directly transferred from ENAGRAM to Instagram when sharing their posts. We have also experienced this particular issue when testing the app ourselves, so we made this limitation explicit from the beginning (i.e., at the briefing). Still, some participants seem to have missed that point and thought it was an unexpected glitch in the software. Problems were also encountered when users attempted to create Instagram stories. Many of them said that their pictures were not properly forwarded to the Instagram app and ended up having multiple publications of the same kind. We did not experience such an issue ourselves during testing but it may be the reason why we observed duplicated records inside the \texttt{EventsDB}. In line with this, some participants reported delays after clicking on ``SHARE!'' or ``POST'' forcing them to click more than once. This may have also contributed to the duplication of entries inside the database and should be then addressed in future ENAGRAM releases.









\section{Study Limitations} \label{sec:limitations}

The adoption of crowdsourcing platforms has become widespread in privacy and security research as they facilitate (to a great extent) the recruitment of study participants and the collection of large amounts of empirical data. Prior work has shown that Prolific samples provide good quality data for conducting survey research on usable privacy and security \cite{tang2022well}. Such is also the case for longitudinal experiments like ours carried out over several days or weeks \cite{kothe2019retention}. Nevertheless, conducting out-of-the-lab experiments also entails a loss of control over participants, giving room to certain types of dishonest practices. For instance, some may claim to meet the eligibility criteria for taking part in the experiment when, in fact, they do not; or may even lack extrinsic motivation (i.e., due to the absence of peer pressure) for completing their tasks \cite{gagne2021run}. Moreover, because of the limited environmental control, online study subjects are prone to get distracted and thus compromise the quality of their answers.

To minimize the effect of these perils we introduced some quality controls throughout the experiment. In particular we included attention questions in all survey instruments and placed a registration code in the app that helped us to assess participants' engagement at the beginning of the study. As described in Section~\ref{expe_design}, we also included an \textit{assessment} stage by the end of the experiment to exclude those who did not use ENAGRAM from completing the final survey. Such an assessment does not offer any guarantee as it relies on participants' self-reports. However, after inspecting the data collected in the \texttt{EventsDB}, we observed that subjects who reported having used the app did use it at least once. Hence, we believe that the control question introduced in the \textit{assessment} stage is a good practice in this type of experiments as it can help in the early detection of loose participants.

The pre-screening of study subjects is often regarded as a best practice when conducting online studies \cite{salminen2021suggestions}. Hence, we used Prolific's built-in qualification features to recruit participants based on their gender (i.e., to ensure balance within study groups), social media usage (Instagram), and language skills (German). In addition, we targeted users who already took part in at least 10 other studies as these are usually more committed and less likely to drop from experiments \cite{salminen2021suggestions}. Last but not least, we also tried to keep the length of both the individual surveys and the full experiment as short as possible to reduce participants' fatigue and minimize the chances of attrition.

As mentioned in Section~\ref{sec:discussion}, the results of this paper are preliminary and aim to give a first impression of ENAGRAM's evaluation features. Still, limitations related to the size and composition of the study sample should be acknowledged. Particularly, we have analyzed a relatively small sample composed exclusively of German-speaking participants. All in all, this means that the results yielded in our study are not representative of the population under analysis. Furthermore, small- to medium-size effects cannot be reliably identified in such a sample and call for future studies with a larger number of participants. One should also note that German-speaking countries (e.g., Germany, Austria, and Switzerland) are typically Western, Educated, Industrialized, Rich, and Democratic (WEIRD) nations  \cite{henrich2010weirdest}. Hence, our sample (and so the study results) are not representative of other populations with different socioeconomic and demographic characteristics. Future work should not only seek to investigate the effects of preventative nudges on larger samples, but also take into account that non-WEIRD individuals (as pointed by \mbox{\citet{dev2020lessons}}) may exhibit different privacy behaviors and concerns. Moreover, prior research has also stressed-out that most studies on digital privacy are based typically on WEIRD samples providing a very narrow view on these matters \cite{alfnes2022your}. Therefore, there is a call for cross-cultural studies that help us better understand the self-disclosure practices, preferences, and concerns across WEIRD and non-WEIRD populations. 

\section{Conclusion} \label{sec:conclusion}

Gathering behavioral evidence on the effectiveness of preventative nudges has become a struggle for privacy and security researchers. The integration barriers imposed by commercial OSNs have limited (to a large extent) the type and amount of empirical data available within the current literature. In turn, nudges targeting online self-disclosure are often evaluated through mock-ups and self-reports but not under realistic conditions. ENAGRAM seeks to aid ongoing investigations on the performance of such nudges by providing a more adequate evaluation environment suitable for conducting longitudinal experiments. The results gathered in our 7-day study show that the app could be leveraged for embedding different nudging strategies and collect insights about their effects on peoples' privacy behavior. Furthermore, the collected data can be aggregated into different performance metrics that, despite being related to the content disclosed by the users, are computed in a privacy-friendly way. That is due to the fact that ENAGRAM only stores hashed versions of the pictures and captions users share, which is adequate to identify behavior changes linked to the presence of nudges (as shown in Section~\ref{sec:behav_in}).

Despite the excitement that nudges arise among privacy researchers, many still see them as a threat to people's autonomy \cite{sunstein2018misconceptions}. That is because nudges often leverage well-known behavioral biases and heuristics to persuade humans toward wiser decisions \cite{zimmermann2021nudge}. Because of the fine line existing between persuasion, manipulation, and coercion, it is not surprising that many have raised concerns over potential unethical uses of nudges. Hence, it is essential to analyze the ethical implications of the approaches tested with ENAGRAM to ensure they do not jeopardize the agency and autonomy of study subjects. \citet{renaud1822} have outlined a set of ethical guidelines applicable to the design of nudges in the context of cybersecurity. We strongly advise those using ENAGRAM as an evaluation framework to adopt these (and other guidelines alike) to ensure their solutions meet ethical requirements from the very beginning.


Throughout this study, we have identified several areas of improvement and directions for future work. One is related to the feedback we received from the participants and the quality of the collected data. As mentioned in Section~\ref{sec:benefits}, some glitches in the current version of ENAGRAM are hindering its usage and may be causing duplicated entries in the \texttt{EventsDB}. Hence, we plan to have a closer look into these issues and apply the corresponding fixes to improve the overall performance of the app. On the other hand, we also aim at embedding off-the shelve machine learning solutions to support the generation of context-aware interventions. This would not only enhance the overall user experience of ENAGRAM but also open new opportunities for the evaluation of preventative nudges. Besides, an assessment with a larger and diverse sample should be conducted over a longer period of time (e.g., 4 weeks) in order to yield more significant and representative results. Particularly, especial attention shall be draw into the cultural differences between WEIRD and non-WEIRD communities as these may considerably impact the effectiveness of preventative nudges in the practice.

Adapting nudges' content and frequency to the individual privacy goals and expectations of each user is still an ongoing research challenge \cite{Bergram2022,peer2020nudge,caraban201923}. The current version of ENAGRAM follows a one-size-fits-all approach in this regard delivering a maximum of 5 interventions per day with a minimum lapse of 60 min. between them. It would be interesting to tailor these parameters to the particular requirements of each user to increase the effectiveness of ENAGRAM's interventions (i.e., of the nudging solution under evaluation). Hence, we also plan to conduct further empirical studies with ENAGRAM to understand the role that frequency and content play in the acceptance of different nudging strategies. Particularly, to determine maximum and minimum intervention thresholds along with personalized strategies for adjusting the content of behavioral interventions.


\begin{acks}
This work was partly supported by Canada's Natural Sciences and Engineering Research Council (NSERC).
\end{acks}

\section*{Supplementary Material}\label{sec:supplementary}

Survey instruments, ENAGRAM software, and study artifacts are available at \url{https://doi.org/10.5281/zenodo.6974704}

\section*{Appendix: Employed constructs}

The reliability of the employed scales was assessed through the Cronbach's Alpha coefficient. In all cases the coefficient was higher than 0.70, which suggests that the items of each construct scale have a relatively high internal consistency (values higher than 0.7 are usually considered ``acceptable''). As mentioned in Section~\ref{expe_design}, all constructs were originally introduced by \citet{Krasnova2009}, except for External Information Privacy Concerns (EIPC) which was elaborated by \citet{morlok2016sharing}. It should be noted that the Perceived Benefits (BEN) construct encompasses \textit{Convenience} (CON), \textit{Relationship Building} (RB), \textit{Self-Representation} (SR), and \textit{Enjoyment} (EN) (we have aggregated these benefits into a single BEN score).
\vspace{1ex}

\textbf{Perceived Benefits (BEN)}
\begin{itemize}
\item CON1: OSNs are convenient to inform all my friends about my ongoing activities.
\item CON2: OSNs allow me to save time when I want to share something new with my friends.
\item CON3: I find OSNs efficient in sharing information with my friends.
\item RB1: Through OSNs I get connected to new people who share my interests.
\item RB2: OSNs helps me to expand my network.
\item RB3: I get to know new people through OSNs.
\item SR1: I try to make a good impression on others on OSNs.
\item SR2: I try to present myself in a favorable way on OSNs.
\item EN1: When I am bored I often log-in to OSNs.
\item EN2: I find OSNs entertaining.
\item EN3: I spend enjoyable and relaxing time on OSNs.
\end{itemize}
\newpage
\textbf{Perceived Privacy Risks (RSK)}
\begin{itemize}
\item RSK1: Overall, I see no real threat to my privacy due to my presence on the OSN (\textit{Reversed}).
\item RSK2: I feel safe publishing my personal information on the OSN (\textit{Reversed}).
\item RSK3: Please rate your overall perception of privacy risk involved when using the OSN (\textit{very safe – very risky}).
\end{itemize}

\textbf{Perceived Control (CTRL)}
\begin{itemize}
\item PC1: I feel in control over the information I provide on the OSN.
\item PC2: Privacy settings allow me to have full control over the information I provide on the OSN.
\item PC3: I feel in control of who can view my information on the OSN.
\end{itemize}

\textbf{External Information Privacy Concerns (EIPC)}
\begin{itemize}
\item EIPC1: It usually bothers me to share pictures of my friends on OSNs.
\item EIPC2: I am concerned that OSNs collect too many pictures of my friends.
\item EIPC3: I am concerned that unauthorized people may access the pictures of my friends that I shared on OSNs.
\item EIPC4: I am concerned that the pictures I share on OSNs may be kept in a non-accurate manner.
\item EIPC5: I am concerned that OSNs may use the pictures of my friends I shared for other purposes without notifying me or getting my authorization.
\item EIPC6: I am concerned that OSNs may sell friends’ pictures I shared to other companies.
\end{itemize}

\bibliographystyle{ACM-Reference-Format}
\bibliography{references}

\end{document}